\documentclass[10pt,twocolumn,twoside]{IEEEtran}

\usepackage{hyperref}
\usepackage{mathtools}
\usepackage{tcolorbox}
\usepackage{color}
\usepackage{theorem}
\usepackage{times,amsmath,epsfig}
\usepackage{amssymb}
\usepackage{cite}
\usepackage[numbers,sort&compress]{natbib}
\def\BibTeX{{\rm B\kern-.05em{\sc i\kern-.025em b}\kern-.08em
T\kern-.1667em\lower.7ex\hbox{E}\kern-.125emX}}
\usepackage{cases}
\usepackage{siunitx}
\usepackage{lipsum}
\usepackage{graphicx}
\usepackage{adjustbox}
\usepackage{array}
\usepackage{multirow}
\usepackage{subcaption}
\usepackage[style=base, font=footnotesize, skip=2pt]{caption}
\let\labelindent\relax
\usepackage[shortlabels]{enumitem}
\usepackage[normalem]{ulem}

\usepackage{algorithmic}
\usepackage{algorithm}

\input{mysymbol.sty}

\newcommand\Tstrut{\rule{0pt}{2ex}}         
\newcommand\Bstrut{\rule[-0.9ex]{0pt}{0pt}}   

\newtheorem{Problem}{Problem}

\newtcolorbox{myblockt}[1]{colback=urblue!5!white,
	colframe=urblue,fonttitle=\bfseries,
	title=#1}

\newtcolorbox{myblock}{colback=urblue!5!white,
	colframe=urblue,fonttitle=\bfseries}

\title{Deep Demixing: Reconstructing the Evolution of Network Epidemics}

\author{
\IEEEauthorblockN{Boning Li$^*$, Gojko Čutura$^*$, Ananthram Swami, and Santiago Segarra}
\thanks{
$^*$Equal contribution. B. Li and S. Segarra are with Rice University, USA. G. Čutura is with  École polytechnique fédérale de Lausanne (EPFL).
A. Swami is with the US Army Research Laboratory, USA. Research was sponsored by the Army Research Office and was accomplished under Cooperative Agreement Number W911NF-19-2-0269. 
The views and conclusions contained in this document are those of the authors and should not be interpreted as representing the official policies, either expressed or implied, of the Army Research Office or the U.S. Government. 
The U.S. Government is authorized to reproduce and distribute reprints for Government purposes notwithstanding any copyright notation herein.
Emails: \{boning.li, segarra\}@rice.edu, gojko.cutura@epfl.ch, ananthram.swami.civ@army.mil.
Preliminary results were presented in~\cite{vcutura2021deep}.
}}

\begin{document}
\maketitle

\begin{abstract}
We propose the deep demixing (DDmix) model, a graph autoencoder that can reconstruct epidemics evolving over networks from partial or aggregated temporal information.
Assuming knowledge of the network topology but not of the epidemic model, our goal is to estimate the complete propagation path of a disease spread.
A data-driven approach is leveraged to overcome the lack of model awareness.
To solve this inverse problem, DDmix is proposed as a graph conditional variational autoencoder that is trained from past epidemic spreads.
DDmix seeks to capture key aspects of the underlying (unknown) spreading dynamics in its latent space.
Using epidemic spreads simulated in synthetic and real-world networks, we demonstrate the accuracy of DDmix by comparing it with multiple (non-graph-aware) learning algorithms.
The generalizability of DDmix is highlighted across different types of networks.
Finally, we showcase that a simple post-processing extension of our proposed method can help identify super-spreaders in the reconstructed propagation path.\looseness=-1
 \end{abstract}


%
\section{Introduction}
\label{S:Introduction}

As the world has gradually restored from the COVID-19 pandemic, accurate reconstruction of past epidemic spreads in communities will assist doctors and policy-makers with understanding the disease better and reducing the risk of future breakouts. 
To date, a lot of stochastic models have been developed to simulate epidemic spreads in networks, including the Susceptible–Infectious-Recovered (SIR) epidemic model~\cite{tornatore2005stability,cooper2020sir,vega2022simlr} and numerous extensions considering additional factors and random perturbations such as vaccination~\cite{tornatore2014sivr}, condition for extinction~\cite{lahrouz2013extinction}, saturated incidence~\cite{zhao2014threshold}, and media coverage~\cite{liu2013sirs}.
Since the basic mechanisms of SARS-CoV-2 (and a lot of other contagious diseases) transmission are through person-to-person contact (e.g., respiratory droplets or via indirect contact)~\cite{el2021sars}, graphs can be useful tools to represent such complex interconnected networks and data defined on them.
Classical signal processing and machine learning methods have recently been extended to encompass this type of graph data~\cite{ortega2018graph,zhang2019general}.
These novel tools have shown promising performance in many popular network science problems such as node classification~\cite{kipf2016semi_supervised}, link prediction~\cite{zhang2018link}, and, more closely related to our case, inference tasks regarding partially observed network processes~\cite{segarra2017blind, zhu2020estimating}.
Indeed, graphs provide versatile modeling tools where nodes might represent anything from single individuals~\cite{zheng2019node_based} to large cities~\cite{pujari2020multi_city} and edges can encode mechanisms of disease propagation (or malware spread). 
While classical graph features such as node centrality and connectivity measures inform public health measures, e.g., immunization strategies and lockdown procedures~\cite{segarra_centrality_2016,cliff2019network, block2020social}, recent research has focused on using more advanced graph learning approaches to better {model and understand epidemics on both individual and community scales}~\cite{tomy2022estimating,la2020epidemiological,derr2020epidemic,shah2020finding,yu2023spatio}.

In this paper, we investigate the particularly challenging use case of reconstructing the entire evolution of an epidemic given partial or aggregated temporal information of the spread.
{In essence, our goal is to estimate the daily health status of each person over a specific time frame, leveraging their interpersonal contact network.}
This is an interesting but inherently ill-posed inverse problem~\cite{kabanikhin2011inverse}.
With precise knowledge of the network process, one may rely on structural features of the initial condition to solve this underdeterminacy~\cite{segarra2017blind}.
Without such knowledge (or when the process is very complex, as in most real-world cases), however, one may resort to model-inspired data-driven solutions.
To this end, we propose a novel neural network architecture, which incorporates a conditional variational autoencoder (CVAE)~\cite{sohn2015learning} and variants of graph neural networks (GNNs)~\cite{kipf2016semi_supervised}.
Our proposed model can be trained to solve the inverse problem from available and/or synthetic data through architecture and training loss inspired by the locality of person-to-person propagation mechanisms.

\vspace{2mm}
\noindent{\bf Related work.}
Temporal epidemic reconstruction has been a central topic among computational epidemiology studies~\cite{rozenshtein2016reconstructing}. 
Depending on the observation model, the reconstruction can be formulated as different well-studied problems.
Given a graph snapshot of all nodes at some moment,~\cite{prakash2012spotting} recovered multiple source nodes that may have caused the infection state based on the Susceptible-Infected (SI) model.
Assuming observation of a subset of active nodes and their activation time,~\cite{xiao2018reconstructing} reconstructed the cascade by formulating and solving a Steiner tree problem.
In general, this body of work largely assumes  precise knowledge of the epidemic model and/or the time of contagion of nodes.
However, due to the complex transmission mechanisms and the incubation period, in reality, such critical information may be inaccurate or even unavailable.
{To overcome these assumptions, we adopt a different approach and rely exclusively on historical data to develop a solution for the reconstruction problem.
To clarify, our work resembles the blind demixing of graph signals, as discussed in related works such as \cite{iglesias2018demixing, iglesias2020blind}. 
However, these prior studies were limited to modeling \emph{linear} network processes, which are inadequate for accurately representing epidemics,} thus prompting the need for \emph{nonlinear} methods like DDmix, the one derived here.

Related to our proposed approach, CVAEs have long been used as effective generative models for a wide range of applications, including trajectory prediction~\cite{feng2019vehicle,ivanovic2020multimodal}, feature recovery~\cite{lopez2017conditional}, and data imputation~\cite{xu2021research}. 
As an extension to variational autoencoders (VAE) on which one cannot control the data generation process (since the latent space is randomly sampled)~\cite{kingma2013auto}, CVAEs are able to generate specific types of data by including a condition (of the data type) on both the encoder and decoder inputs~\cite{sohn2015learning}.
Motivated by this success,~\cite{balakrishnan2019visual} proposed a CVAE based on convolutional neural networks (CNNs) for reconstructing video from temporally collapsed images.  
Moreover, these tools have been extended to the non-Euclidean domain of irregularly structured data that are best modeled as graphs~\cite{kipf2016variational} with graph model structures as simple as graph convolutional networks (GCN)~\cite{kipf2016semi_supervised}.
Previously, most applications have been focused on molecular design~\cite{lim2018molecular,lee2022mgcvae} or node embedding procedures~\cite{hamilton2017inductive}. 
{
More closely related to our interest, Chen et al.~\cite{ling2022source} trained a graph VAE for source localization, but this task is only one piece of a larger puzzle.
In the context of broader disease tracking goals, our objective of reconstructing the complete outbreak not only treats source localization as a natural byproduct but could also provide a more comprehensive understanding of how the disease is spreading and evolving over time. 
This could help inform more effective strategies for prevention, containment, and treatment. 
Another limitation of~\cite{ling2022source} is the use of basic spread models like SI and SIR as opposed to more realistic settings such as SIRS and SIRSD.
}

In this paper, we develop a version of a CVAE taking GNNs as its backbone.
More precisely, the encoder, decoder, prior, and posterior networks of our proposed architecture all bear a resemblance to the Graph \mbox{U-Net}~\cite{gao2019graph} style.
To the best of our knowledge, this is the first implementation of a graph CVAE for the study of epidemics.

\vspace{2mm}
\noindent{\bf Contribution.} 
The contributions of our paper are threefold:
\begin{enumerate}[wide,label=\roman*),topsep=0pt,itemsep=-1ex,partopsep=1ex,parsep=1ex,labelindent=0pt]
    \item We propose DDmix, a novel graph CVAE architecture for the temporal reconstruction of partially-observed network processes; we further design a locality regularizer to improve its performance in epidemics.
    \item We show the successful implementation of DDmix to infer the evolution of epidemics surpassing non-graph-aware deep learning methods in accuracy and generalizability.
    \item We explore simple (no need to retrain) but effective extensions of DDmix for tracing the source of and assessing super-spreaders in an epidemic.
\end{enumerate}

\section{System Model and Problem Formulation}
\label{S:problem}

In Section~\ref{ss:nm}, we introduce the network model as the basic person-to-person contact network. 
In Section~\ref{ss:em}, we briefly review the epidemic model for synthetic data generation. 
The formal problem statement is given in Section~\ref{ss:ps}, followed by a few remarks on our goal and approach.\looseness=-1

\subsection{Network model}
\label{ss:nm}

Let $\ccalG{\,=\,}(\ccalV, \ccalE)$ denote an undirected graph where $\ccalV{\,=\,}[N]{\,=\,}\{1,2,\cdots,N\}$ is the vertex (or node) set and $\ccalE$ represents the set of edges. 
The existence of an edge $(i,j){\,\in\,}\ccalE$ between nodes $i$ and $j$ corresponds to a non-zero entry $A_{ij}$ in the adjacency matrix $\bbA{\,\in\,}\mbR^{N\times N}$.
For undirected and unweighted graphs, $A_{ij}{\,=\,}A_{ji}{\,=\,}1$ for all $(i,j){\,\in\,}\ccalE$, and $A_{ij}{\,=\,}A_{ji}{\,=\,}0$ otherwise, thus fully characterizing the graph structure.\looseness=-1 

To explain the physical interpretation, a node represents an individual and an edge encodes the possibility of contagion between a pair of individuals. In this way, we can model our interconnected system as a graph and the transmission of an epidemic as information flow in this graph. 
More precisely, we model the state of the nodes at any given time using graph signals, i.e., maps $x: \ccalV{\,\to\,}\reals$ from the node set into the reals. 
For convenience, we use the vector representation $\bbx{\,\in\,}\reals_{+}^N$ with the subscript $(\cdot)_+$ denoting elementwise nonnegativity, in which $x_i$ collects the value of the graph signal at node $i$.
We also consider (binary) time series, where $\{\bby^{(t)}\}_{t=1}^T$ contains ordered binary graph signal vectors $\bby^{(t)}{\,\in\,}\mbZ_2^N$. 
To be exact, $y^{(t)}_i{\,=\,}1$ indicates that node $i$ was in the infected state at the time instant $t$ while $y^{(t)}_i{\,=\,}0$ indicates that it was not infected at that time.

\subsection{Epidemic model}
\label{ss:em}

We mainly focus on the well-established SIRS model~\cite{zheng2019node_based}.
Nevertheless, we underscore that \emph{no knowledge} of the underlying epidemic model is assumed by our proposed solution in reconstructing the epidemic spread.

\begin{figure}
    \centering
    \begin{subfigure}[b]{.27\linewidth}
        \centering
        \includegraphics[trim=.2cm .5cm .2cm .5cm, width=.9\linewidth]{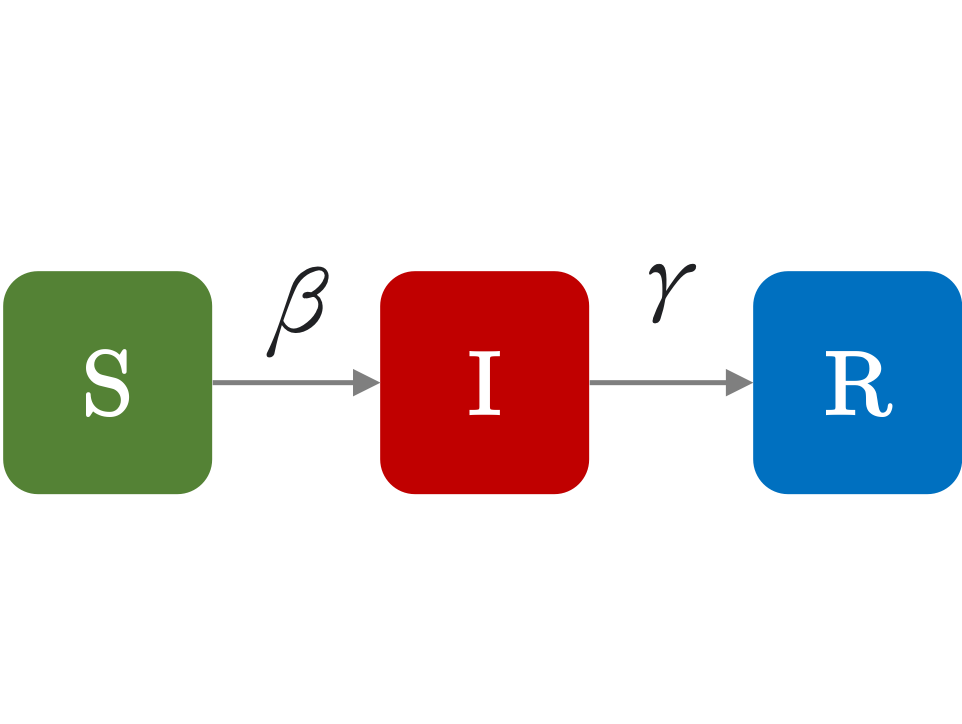}
        \caption{SIR}\label{ff:sir}
    \end{subfigure}\hfill
    \begin{subfigure}[b]{.27\linewidth}
        \centering
        \includegraphics[trim=.2cm .5cm .2cm .5cm, width=.9\linewidth]{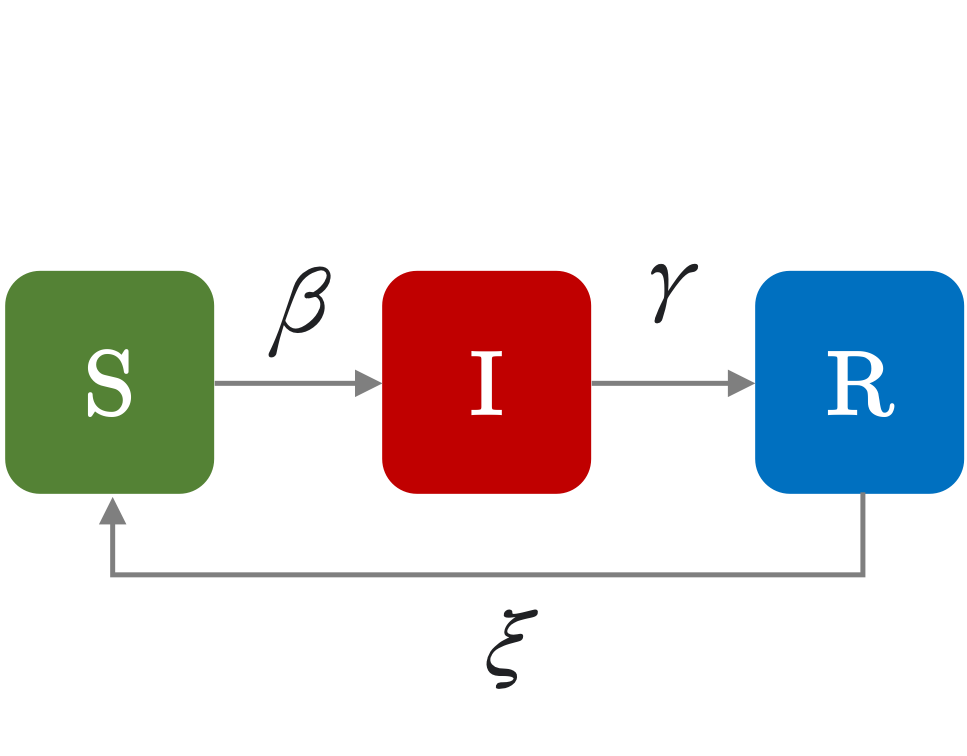}
        \caption{SIRS}\label{ff:sirs}
    \end{subfigure}\hfill
    \begin{subfigure}[b]{.27\linewidth}
        \centering
        \includegraphics[trim=.2cm .5cm .2cm .5cm, width=.9\linewidth]{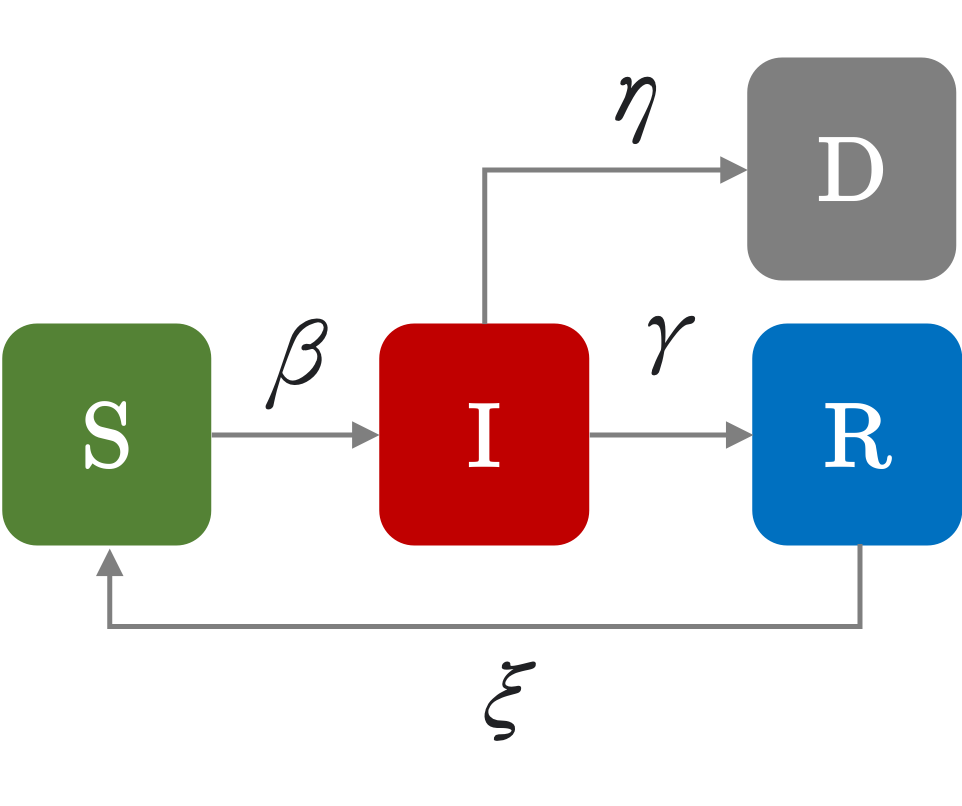}
        \caption{SIRSD}\label{ff:sirsd}
    \end{subfigure}
    \caption{Illustrations of three epidemic models. The four possible states are S for susceptible, I for infectious, R for recovered, and D for dead.
    Nodes in the I state are considered to be having the disease, or with signal 1. 
    In any other state, the node signal is 0.
    Connections from one state to another are labeled with the corresponding transition probability. 
    }\label{ff:ep_model}
\end{figure}

The classical SIR epidemic model~(Fig.~\ref{ff:sir}) assumes that at every discrete time step $t$,
i)~a healthy yet susceptible node $i$ can become infected (and infectious) by any of its infectious neighbors in $\ccalG$ with a probability $\beta$, independent of other nodes, and 
ii)~an infected node can recover from (and be immune to) the disease with probability $\gamma$. 
While SIR assumes lifelong immunity, the extended SIRS model~(Fig.~\ref{ff:sirs}) considers 
iii)~an additional transition from the recovered state back to the susceptible state with probability $\xi$, which can happen due to the drop in immunity over time or the emergence of new variants of the virus. 
Furthermore, if we consider death as a possible outcome of infection, i.e., iv)~an infected node may transit to and permanently stay in a new state with probability $\eta$, we get the SIRSD model (Fig.~\ref{ff:sirsd}). 
In the rest of this paper, we primarily focus on the SIRS model unless explicitly stated otherwise.

For an arbitrary epidemic spread $\bbY{\,:=\,}\{\bby^{(t)}\}_{t=1}^T$ of interest, we assume that the available information is partial and given by a single graph signal $\bbx{\,=\,}f_m(\{\bby^{(t)}\}_{t=1}^T)$ resulting from some mixing function $f_m:\mbR^{N\times T}{\,\to\,}\mbR^{N}$.
Here we focus on the case where $\bbx$ represents an average aggregation of the recorded infections, i.e.,
\begin{equation}\label{E:aggrregation}
    \bbx{\,=\,}\frac{1}{T} \sum_{t=1}^{T} \bby^{(t)}.
\end{equation}
Intuitively, we get to observe if a person was ever infected or not over a given period of time as well as the length of the infection (either self-reported or estimated by proxies such as serology tests), but we do not have precise information on when the infection started and ended. 

\subsection{Problem statement}
\label{ss:ps}

With the notation introduced in Sections~\ref{ss:nm} and~\ref{ss:em}, we formally state our problem as follows.

\begin{Problem}\label{P:main}
    Consider a known graph of fixed structure $\ccalG{\,=\,}(\ccalV, \ccalE)$ where $|\ccalV|{\,=\,}N$.
    An unknown epidemic spread {modeled by fixed yet unknown parameters (Fig.~\ref{ff:ep_model})} $\bbY{\,=\,}\{\bby^{(t)}\}_{t=1}^T{\,\in\,}\mbZ_2^{N\times T}$ evolves over $\ccalG$ for a total of $T{\,>\,}1$ time steps. 
    Observing the temporally mixed signal $\bbx{\,\in\,}\mbR^{N}$ as in~\eqref{E:aggrregation}, estimate the complete evolution of the spread as $\hbY{\,=\,}\{\hat{\bby}^{(t)}\}_{t=1}^T$.
\end{Problem}

Before presenting our proposed solution to Problem~\ref{P:main}, a few comments are in order.
First, the observation model in~\eqref{E:aggrregation} can be interpreted as encoding vague self-reported information such as `I tested positive for five days over last summer, though I don't remember the exact dates'.
Second, given the (temporal) aggregation of many signals, Problem~\ref{P:main} aims to exploit the graph structure to separate them apart, or to \emph{demix} the graph signals.
Even considering the simpler setting of linear network processes, graph signal demixing is already a challenging and highly underdetermined problem.
Finally, most existing methods assume full or partial knowledge of the epidemic model to solve this ill-posed and nonlinear problem~\cite{prakash2012spotting, lappas2010finding}.
By contrast, we propose a data-driven approach where we do not rely on this idealistic assumption. 
More precisely, we assume no knowledge of the underlying epidemic model except that the process is driven by the topology of $\ccalG$.
Using past known observation pairs $(\bbx, \{\bby^{(t)}\}_{t=1}^T)$, we can train our model to demix new signals $\bbx_{\text{test}}$.
Note that the training data are easy to acquire, for we can simulate as many random realizations of the process as we need, as discussed in Section~\ref{ss:em}.
Another advantage is that no manual annotation is involved in constructing the training dataset.



\section{DDmix method}
\label{S:Graph_CVAE}

We now present DDmix in detail.
Section~\ref{ss:gcva} introduces the CVAE framework as a solution to Problem~\ref{P:main} and Section~\ref{ss:ddmix-arch} outlines the proposed modular architecture of DDmix.

\subsection{Graph CVAE for temporal demixing}
\label{ss:gcva}

Let us begin by assuming access to $M$ pairs of i.i.d. observations and ground-truth spreading information $\{(\bbx_j, \bbY_j)\}_{j=1}^{M}$ where $\bbx{\,\in\,}\mbR_{+}^{N}$ represents an observed epidemic signal and $\bbY{\,\in\,}\mbZ_2^{N{\times}T}$ its corresponding true spread.
We aim to leverage the available observations to solve Problem~\ref{P:main} by estimating the conditional distribution $p(\bbY|\bbx)$ for the generation (or reconstruction) process of interest.
When this distribution is obtained, given any observation $\bbx_{M+1}$ that is likely never seen in the past, we can sample our candidate solution $\bbY_{M+1}$ to Problem~\ref{P:main} from $p(\bbY|\bbx_{M+1})$.
In determining the distribution of interest $p(\bbY|\bbx)$, we adopt a CVAE probabilistic model~\cite{sohn2015learning} with a latent variable $\bbz$ that seeks to model features of the temporal variation of the epidemic collapsed in $\bbx$. 
Intuitively, we want $\bbz$ to capture key characteristics of the underlying (unknown) spreading dynamics.

We model $\bbz$ as conditionally Gaussian given $\bbx$, i.e.
\begin{alignat}{3}
    p_{\phi}(\bbz{\,|\,}\bbx)&{\,=\,}\mathcal{N}(\mu_{\phi}(\bbx),\,\sigma^2_{\phi}(\bbx)),\label{E:distribution_phi}
\end{alignat}
where we have made explicit the fact that the mean and variance of the distribution of $\bbz$ are functions of $\bbx$ parametrized by $\phi$.
We use a specific neural network construction (see Section~\ref{ss:ddmix-arch}) as our parameterization scheme and we denote $p_{\phi}(\bbz{\,|\,}\bbx)$ as our \emph{prior network}.
Following the CVAE framework, we define the distribution of our variable of interest as
\begin{alignat}{3}
    p_{\theta}(\bbY{\,|\,}\bbx, \bbz)&{\,=\,}\ccalN(g_{\theta}(\bbx, \bbz),\,\sigma_{y}^{2}\,\bbI\,).\label{E:distribution_theta}
\end{alignat}
Under this probabilistic model, a parametric deprojection function $g_\theta$ models the expected value of the temporal evolution $\bbY$ as a function of our observation $\bbx$ and the corresponding true latent variable $\bbz$.
A common noise variance $\sigma_{y}^{2}$ is assumed for all entries.
We denote this parametric function as our \emph{deprojection network}.

From the two conditional probabilities introduced above, it follows that we
can compute the distribution of interest as
\begin{equation*}\label{E:distribution_theta_phi}
    p_{\theta, \phi}(\bbY{\,|\,}\bbx){\,=\,}\int_{\bbz} p_{\theta}(\bbY{\,|\,}\bbx, \bbz)\,\,p_{\phi}(\bbz{\,|\,}\bbx) \,d\bbz,
\end{equation*}
which involves, however, an intractable integral to solve over $\bbz$ even for fairly simple parametrizations of $p_{\theta}$ and $p_{\phi}$.
In general, it is thus a challenging endeavor to determine the parameters $\theta$ and $\phi$ that maximize the likelihood of the observed $M$ pairs $(\bbx, \bbY)$ in closed form.

\begin{figure*}[t]
	\centering
    \includegraphics[width=\linewidth]{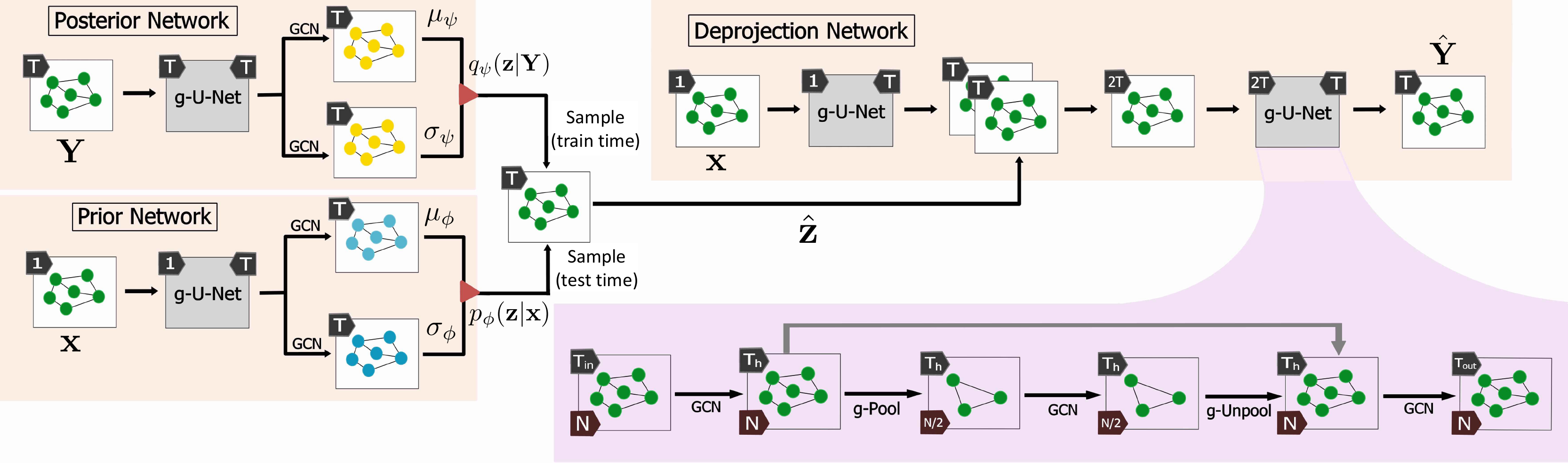}
    \vspace{1.5mm}
	\caption{
     Overall view of our proposed deep demixing (DDmix) architecture based on graph neural networks for the temporal reconstruction of epidemics. 
     The three subnetworks -- prior, posterior, and deprojection -- parameterize key functions in the assumed probabilistic model.
 }\label{fig:architecture}
\end{figure*}

Instead, we follow the well-established route of variational inference and propose a tractable loss inspired by the evidence lower bound~\cite{blei2017variational} that can be optimized via stochastic gradient descent. 
To achieve that, we introduce a parametrization for the posterior probability of $\bbz$ given $\bbY$
\begin{equation}\label{E:distribution_psi}
    q_{\psi}(\bbz{\,|\,}\bbY){\,=\,}\ccalN(\mu_{\psi}(\bbY), \sigma^2_{\psi}(\bbY)).
\end{equation}
This constitutes the third and last parametric component in our architecture, namely the \emph{posterior network}; see Fig.~\ref{fig:architecture}.

\vspace{2mm}

\noindent{\bf Loss function.} To train our three networks (prior, posterior, and deprojection), we consider the following compound loss
\begin{align}\label{E:loss}
        \ccalL_{\bbtheta, \bbphi, \psi}(\bbx, \bbY){\,=\,}& \, \eta_1 L_1( p_{\phi}(\bbz |\bbx), q_{\psi}(\bbz|\bbY) ){\,+\,}\eta_2 L_2(g_{\theta}(\bbx, \hbz), \bbY)\nonumber \\       
        & {\,+\,}\eta_3 R_1(\bbtheta, \bbphi, \bbpsi){\,+\,}\eta_4 R_2(g_{\theta}(\bbx, \hbz)),
\end{align}
consisting of four additive terms -- two fitting terms $L_1 (\cdot)$ and $L_2(\cdot)$ and two regularization terms $R_1 (\cdot)$ and $R_2(\cdot)$. 
Each term is weighted by its corresponding scalar weight from the set $\{\eta_i\}_{i=1}^4$. 
In the remainder of this section, we will explain our implementation of each term in detail.

We define the first fitting loss term as the KL divergence between the prior and posterior distributions
\begin{equation}\label{E:loss_1}
        L_1(p_{\phi}(\bbz |\bbx), q_{\psi}(\bbz|\bbY)){\,=\,}D_{\text{KL}} (p_{\phi}(\bbz |\bbx){\,\|\,}q_{\psi}(\bbz|\bbY)). 
\end{equation}
The purpose is that, during testing, the samples that we draw from the prior distribution would resemble the samples drawn from the (more informative) posterior distribution.
In this way, during testing, when $\bbY$ is not available and, thus, we cannot rely on the posterior network, the latent variable $\bbz$ drawn from the prior network $p_{\phi}(\bbz |\bbx)$ would still constitute a reasonable input to the deprojection network.

The second fitting term is the reconstruction loss measuring the distance between the inferred and the true time series
\begin{equation}\label{E:loss_2}
        L_2(\hbY, \bbY){\,=\,}- \bbY \log \hbY - (1-\bbY) \log(1 - \hbY),
\end{equation}
where $\hbY{\,=\,}g_{\theta}(\bbx, \hbz)$ is the reconstruction, with the sampled $\hbz{\,\sim\,}\ccalN(\mu_\psi, \sigma^2_\psi)$ given by the posterior network.
In defining this term, notice that the entries of $\bbY$ are binary (either infected or not); thus, we optimize for a binary cross entropy loss over the trainable deprojection and posterior parameters $(\bbtheta, \bbpsi)$.\looseness=-1

When it comes to regularization, we first implement a simple $\ell_2$ penalty on all our parameters $(\bbtheta, \bbphi, \bbpsi)$,
\begin{equation}\label{E:Reg_1}
    R_1(\bbtheta, \bbphi, \bbpsi){\,=\,}\sum_{i=1}^{|\bbtheta|}\theta_i^2+\sum_{i=1}^{|\bbphi|}\phi_i^2+\sum_{i=1}^{|\bbpsi|}\psi_i^2,
\end{equation}
which aims to force the magnitude of parameters to be small in order to reduce the chance of overfitting. 

More specific to the problem at hand, in the second regularizer, we incorporate domain knowledge that the epidemic evolution should be local in $\ccalG$.
In other words, the disease can only propagate through the contact graph --
a node has a possibility of being infected at time $t$ (${\,>\,}1$) only if itself or at least one of its direct neighbors was infectious at the previous time $t{\,-\,}1$.
To be precise, we define\looseness=-1
\begin{equation}\label{E:Reg_2}
    R_2(\hbY){\,=\,}\left\| \sum_{t=2}^{T} \left[\hat{\bby}^{(t)} - (\bbA{\,+\,}\bbI) \hat{\bby}^{(t-1)}\right]_{+} \right\|_1,
\end{equation}
where $[\,\cdot\,]_+$ denotes the positive projection and $\hat{\bby}^{(t)}$ is the $t$-th column of $\hbY{\,=\,}g_{\theta}(\bbx, \hbz)$.
The intuition behind $R_2$ is to penalize the appearance of an infected node at time $t$ when neither itself nor any of its neighbors were infected at time $t{\,-\,}1$ in the reconstructed time series.
As in the case of the reconstruction loss, the scope of this term is over the deprojection and posterior parameters $(\bbtheta, \bbpsi)$.

\subsection{DDmix architecture}
\label{ss:ddmix-arch}

So far, we have presented our approach in terms of generic parametric distributions $p_{\phi}(\bbz{\,|\,}\bbx)$, $p_{\theta}(\bbY{\,|\,}\bbx, \bbz)$, and $q_{\psi}(\bbz{\,|\,}\bbY)$.
In this section, we provide explicit forms for the parameterizations used.
Given the graph structure of our data and the nature of the problem being solved, our parameterizations take the form of graph neural networks, thus explicitly relying on the underlying graph.
See Fig.~\ref{fig:architecture} for an overall schematic view of the proposed architecture.

\vspace{2mm}\noindent{\bf Graph convolutional network.} 
We adopt the GCN layer as our primary building block, for its simple structure and permutation equivariance (i.e., the performance of the overall architecture does not depend on the indexing of the nodes in the graph). 
The GCN provides an efficient layer-wise propagation rule that is based on a first-order approximation of spectral convolutions on graphs. 
It achieves this by the forward-propagation method
\begin{equation}\label{E:gcn_propagation}
    \bbH{\,=\,}\sigma(\tilde{\bbA} \bbH_0 \bbW),
\end{equation}
where $\bbH_0{\,\in\,}\reals^{N{\times}d_0}$ is the input to the layer, $\bbH{\,\in\,}\reals^{N{\times}d}$ is its output (likely also the input to the next layer), while $d_0$ and $d$ are some predefined feature dimensions. 
Accordingly, $\bbW^{d_0{\times}d}$ is a matrix of trainable weights for that layer, and $\sigma(\cdot)$ is an elementwise nonlinear activation function such as ReLU.
The normalized adjacency matrix $\tilde{\bbA}{\,\in\,}\reals_{+}^{N{\times}N}$ is given by $\tilde{\bbA}{\,=\,}\hat{\bbD}^{-\frac{1}{2}} (\bbA{\,+\,}\bbI) \hat{\bbD}^{-\frac{1}{2}}$, where $\hat{\bbD}{\,\in\,}\mbZ_{+}^{N{\times}N}$ is the diagonal node degree matrix.\looseness=-1

\vspace{2mm}\noindent{\bf Graph pooling and unpooling.}
Graph pooling (gPool) and graph unpooling (gUnpool) layers, proposed in~\cite{gao2019graph}, respectively present graph-aware down-sampling and up-sampling methods, taking into account and preserving the structure of the graph, as well as the features of its nodes. 
More specifically, the \emph{gPool} layer allows for adaptive selection of a subset of nodes in the input graph, resulting in one of its subgraphs. 
For the purpose of choosing the best subset of nodes, a trainable projection vector $\bbp$ is introduced. 
Using $\bbp$, all nodes of the input graph are projected into a 1-dimensional vector $\bby$, i.e. for node $i$, whose feature vector is given by $\bbh_i$, the 1-dimensional projection of $\bbh_i$ onto $\bbp$ is calculated as $y_i{\,=\,}\bbh_i^\top \frac{\bbp}{\|\bbp\|}$. 
The resulting vector $\bby$ is used to perform $k$-max pooling, selecting the nodes with $k$ largest scalar projections $y_i$, returning a list of node indices
\begin{equation}\label{E:pooling}
    \bba{\,=\,}\text{max}(\bby, k), \,\,\, 0{\,\leq\,}a_j{\,<\,}N,\,\forall\,j{\,=\,}1,...,k.
\end{equation}
The intuition behind this is that each projection $y_i$ effectively measures how much information of node $i$ can be preserved after projecting its features onto $\bbp$.
Since the goal is to choose the most informative nodes, $k$-max pooling is performed.

On the other hand, the \emph{gUnpool} layer performs the inverse operation to the one of gPool, restoring the original input graph structure. 
All the input nodes are placed back into their original positions, which were recorded right before the down-sampling phase. 
Furthermore, the nodes which were selected by the pooling layer (i.e. are the input into the unpooling layer) keep their new feature vectors, whereas the feature vectors of the rest of the nodes are kept as zero. 

\vspace{2mm}\noindent{\bf Graph U-Net.} 
By combining the previously introduced layers (GCN, gPool, gUnpool), one can build the graph U-Net (\mbox{g-U-Net}) model. 
The model takes a graph as input, which first gets passed through an embedding layer in order to map the original node features into a lower-dimensional representation. 
What follows is the encoder-decoder structure, as is the case in the original convolutional U-net architecture~\cite{ronneberger2015u}. 
The encoder is constructed by stacking multiple blocks made of gPool and GCN layers.
On the other side, the decoder has the same number of blocks, but instead of gPool, it uses the up-sampling gUnpool layer to restore the graph to its higher resolution structure; see Fig.~\ref{fig:architecture}. 
The GCN layers in both the encoder and decoder blocks play the role of aggregating node neighborhood information. 
Finally, skip-connections are added between encoder and decoder blocks of the same depth, and they can either be implemented as an addition or concatenation of corresponding feature maps.

\vspace{2mm}\noindent{\bf Prior network $p_{\phi}(\bbz{\,|\,}\bbx)$.} 
The input to the prior network is $\bbx$, an observation of the aggregated graph signal.
Each node's input feature is a scalar value $x_i$ (where $i$ is the node index) which is then passed through the \mbox{g-U-Net} block with the number of features per node increased to $T$. 
Following the \mbox{g-U-Net} block, two parallel GCN layers are then applied to its output, resulting in the mean $\mu_{\phi}$ and STD $\sigma_{\phi}$ of the Gaussian distribution $p_{\phi}$.

\vspace{2mm}\noindent{\bf Posterior network $q_{\psi}(\bbz{\,|\,}\bbY)$.} 
Analogously, we design the posterior network with the only change being the input dimensions.
For the posterior network, the number of features per node in the input signal $\bbY$ is $T$, corresponding to the complete, uncollapsed time series.
Similarly, the posterior network produces the mean {$\mu_{\psi}$} and STD {$\sigma_{\psi}$} of the Gaussian distribution $q_{\psi}$.
With these obtained parameters, we can then sample the latent variable $\hbz$ from the corresponding distribution, a step denoted by the red triangles in Fig.~\ref{fig:architecture}.
This $\hbz$ will directly contribute to the inference of the final reconstruction in the next step, yet it should be clarified that it is sampled using the posterior network $q_{\psi}$ during training while the prior network $p_\phi$ is used during testing.
Understandably, during testing we would not have had the true signal $\bbY$, practically excluding the involvement of $q_{\psi}$, for it takes $\bbY$ as input.

\vspace{2mm}\noindent{\bf Deprojection network $p_{\theta}(\bbY{\,|\,}\bbx, \bbz)$.} 
The deprojection network takes both the observed $\bbx$ and the sampled latent variable $\hbz$ as input [see~\eqref{E:distribution_theta}].
It ultimately seeks to output a good estimate $\hbY{\,=\,}g_\theta(\bbx, \hbz)$ of the temporal evolution that is as close to the true evolution $\bbY$ as possible.
In our implementation, we first pass $\bbx$ through a \mbox{g-U-Net} block while increasing the number of node features from $1$ to $T$, whose output is concatenated with $\hbz$, resulting in a graph signal with $2T$ features per node. 
This intermediate signal is then passed through the last \mbox{g-U-Net} block in DDmix, through which the information in $\bbx$ and $\hbz$ is combined and the number of node features is reduced to $T$ to match the prefixed length of the time series being estimated.

\vspace{1mm}
In summary, the graph-aware variational architecture presented in Fig.~\ref{fig:architecture} can be trained end-to-end given observed epidemic evolutions $\{(\bbx_j, \bbY_j)\}_{j=1}^{M}$ and, once trained, it can be used to solve Problem~\ref{P:main}, i.e., to reconstruct the epidemic $\bbY$ from an observation $\bbx$.
In the next section, we demonstrate the performance of this approach.

\section{Numerical Experiments}\label{S:Experiments}

Through various synthetic topologies and real-world contact graphs, we illustrate the performance of DDmix in diverse settings.\footnote{Code to replicate the numerical experiments here presented can be found at
\href{https://github.com/gojkoc54/Deep_demixing}{https://github.com/gojkoc54/Deep\_demixing}.}
We compare DDmix with the following baselines:

\vspace{2mm}\noindent 
1) \textbf{MLP:} 
A multi-layer perceptron that takes $\bbx$ as input, propagates it through 3 hidden layers with \{$\frac{1}{4}N T$, $\frac{1}{4}N T$, $N T$\} neurons and ReLU activations. 
The output layer of size $N T$ estimates the vertical concatenation of the columns $\bby^{(t)}$ of $\bbY$.

\vspace{1mm}\noindent 
2) \textbf{LSTM:} 
A one-to-many bidirectional long short-term memory (LSTM) recurrent network~\cite{sak2014long} of length $T$ that unrolls $\bbx$ into $T$ $N$-element vectors through 2 hidden layers each containing $2N$ neurons. 
The full epidemic evolution $\bbY$ is estimated by the concatenation of all output vectors.

\vspace{1mm}\noindent 
3) \textbf{CNN-nodes:} 
A convolutional neural network (CNN) that takes $\bbx$ as input and performs one-dimensional depthwise convolutions with kernels of size 3, and both stride and padding equal to 1. 
No pooling is performed and the number of channels gradually increases from $1$ to $T$ ($1, T/4, T/2, T$) so that the $N\times T$ two-dimensional output estimates $\bbY$.

\vspace{1mm}\noindent 
4) \textbf{CNN-time:} 
A CNN that performs one-dimensional transposed convolutions~\cite{dumoulin2016guide} over the temporal dimension with fractional strides.
For each node $i$, CNN-time takes $x_i$ as one-dimensional input and after 6 blocks of convolution and elementwise ReLU activation, it computes a $T$-dimensional output estimating $[y^{(1)}_i, \ldots, y^{(T)}_i]$.

\vspace{2mm}\noindent
{Table~\ref{tab:model-complex} provides an overview of the complexity of these models ($T=20$) by reporting the number of trainable parameters.
From this table, MLP and LSTM are likely to be severely overparametrized due to their fully connected architectures. 
The CNN-based models see a drastic reduction in the number of parameters owing to the local convolutional kernels utilized. 
While models with fewer trainable parameters can generalize better and combat overfitting, the issue of underparameterization may rise, as in CNN-nodes, which can limit the upper bound of learning performance.
Our proposed method, despite having a specialized architecture, essentially extends CNN to the graph domain and thus benefits from a relatively small parameter set as well.
}

\begin{table}[t]
    \captionsetup{justification=centering}
    \caption{
    Comparison of the number of trainable parameters in each model. 
    }\label{tab:model-complex}
    \centering
    \resizebox{\linewidth}{!}{%
        \begin{tabular}{r|ccccc}
            \hline
            Complexity & DDmix & CNN-time & CNN-nodes & LSTM & MLP \\\hline
            \# of parameters & 6640  & 2559  & 44  & 5953K  & 5355K \\\hline
        \end{tabular}
    }
\end{table}

It should also be noted that, although data-driven, the four conventional machine learning baselines considered are graph agnostic. 
More precisely, MLP incorporates fully-connected layers, thus overlooking the notion of locality in $\ccalG$.
LSTM, compared to the former, emphasizes the temporal dependence of output sequences yet still lacks explicit incorporation of locality.
CNN-nodes, relying on convolutional filters, assumes a notion of locality inherited by the indexing of the nodes, which need not align with the true notion of locality driving the underlying epidemic.
Finally, CNN-time ignores the effect of interconnections between nodes and seeks to solve the temporal reconstruction for each node independently. 
By contrast, DDmix explicitly incorporates the graph structure in its architecture, redounding in higher performance and enhancing its generalizability.

\subsection{Reconstruction with varying topological parameters}\label{sec:exp:topo-params}

\noindent
{\bf Initial comparison of demixing methods.}\label{sec:exp:comp-perf}
We use synthetic random geometric (RG) graphs generated by placing $N$ nodes uniformly at random in the unit cube. 
An edge is inserted between two nodes if their Euclidean distance is less than or equal to $d_{r}$. 
The baseline RG graph has $N_0=100$ and $d_{r_0}{\,=\,}0.30$.
Data samples are independent simulations of epidemic evolutions whose full length is $T{\,=\,}20$ days, in addition to which we consider partial evolutions of $T{\,=\,}10$ as the 10 \textit{leading} days.
In each simulation, the epidemic starts at a random node chosen from the uniform distribution and evolves with model parameters (see Section~\ref{S:problem}) set to $\beta{\,=\,}0.5$, $\delta{\,=\,}0.05$, $\gamma{\,=\,}0.005$, and, if applicable, $\mu=0.1$.
We set $\eta_1{\,=\,}\eta_2{\,=\,}\eta_4{\,=\,}1$ and $\eta_3{\,=\,}5\times10^{-4}$ in the loss~\eqref{E:loss}.
The Adam optimizer~\cite{kingma2014adam} is adopted with a fixed learning rate of $0.01$, $\beta_1{\,=\,}0.9$ and $\beta_2{\,=\,}0.999$.\looseness=-1

Next, we introduce the training and evaluation protocols as follows.
Training takes place in mini-batches of size 4 for a maximum of 50 epochs (early stop in case the validation loss stops dropping for 10 epochs). 
Unless otherwise stated, we train and validate the models with 100 and 1000 random simulations in the same graph and report final results based on another 1000 random simulations on a different test graph drawn from the same random graph model. 
The number of training samples is chosen guided by our preliminary work~\cite{vcutura2021deep}, where we found that this number could be reduced from 4500 to 100 without harming the learning performance.
Three random runs of 3-fold cross-validation are performed and the average performance is taken to ensure the validity of our results.
It is worth reiterating that DDmix is agnostic to the specific epidemic model used to generate the spread data, and as such, does not require any knowledge of the underlying model's parameters.

\begin{figure}[t]
    \begin{center}
        \includegraphics[clip,width=\linewidth]{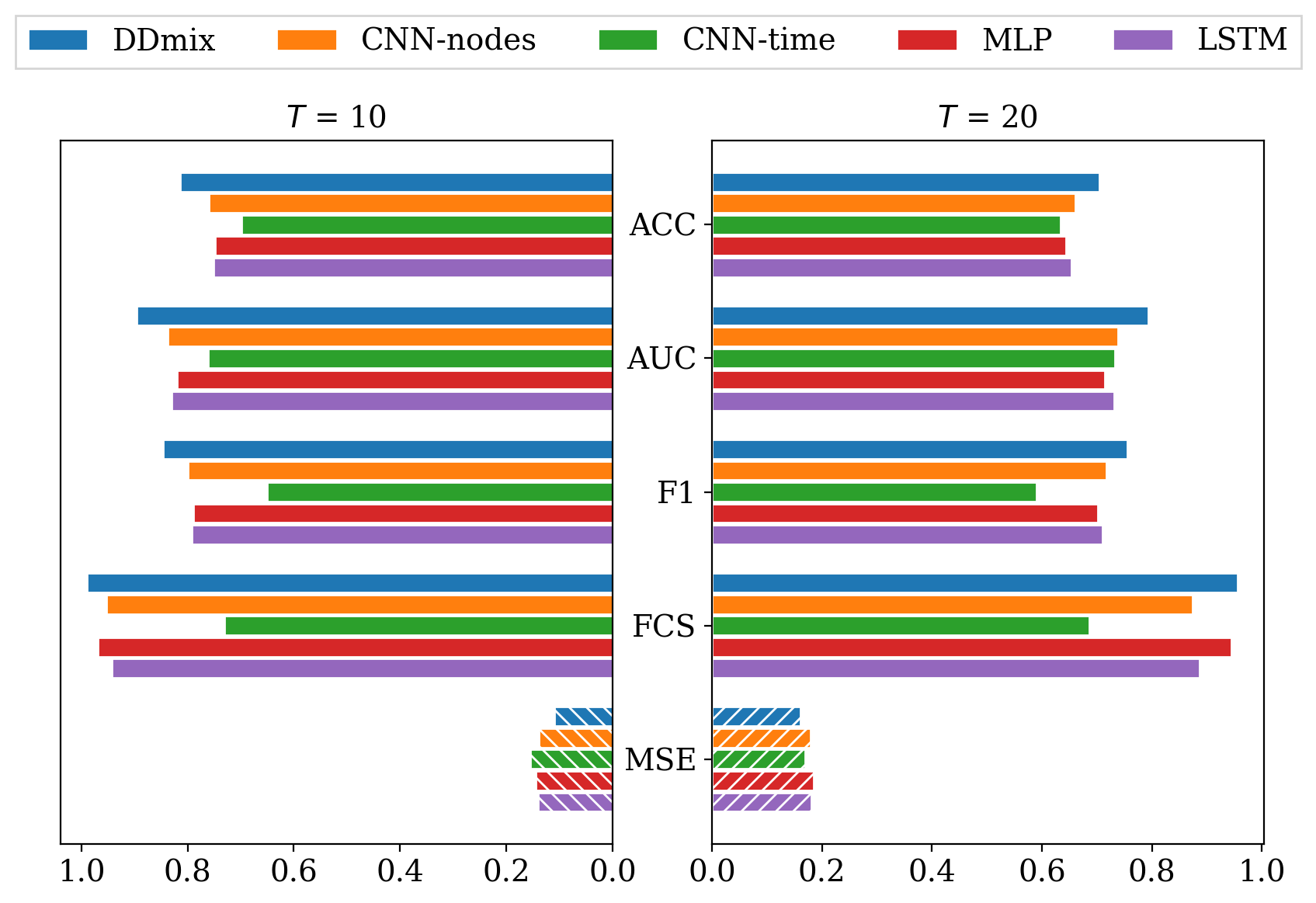}
        \caption{
        Comparison of models by their temporal reconstruction performance of SIRS data, including the high-is-better metrics of prediction accuracy, AUC, F1, and FCS scores, as well as the lower-is-better MSE. 
        }
        \label{fig:performance_comparison}
    \end{center}
\end{figure}

Figure~\ref{fig:performance_comparison} illustrates four correctness metrics for temporal reconstruction performance, namely 
1)~ACC: the straightforward classification accuracy defined as $\frac{1}{NT} |\bbY - \bar{\bbY}|$ where $\bar{\bbY}$ is the binary prediction of $\bbY$;
2)~AUC: measures the area underneath the ROC curve;
3)~F1: the harmonic mean of precision and recall;
4)~FCS: fractional cosine similarity (FCS) is defined as the cosine similarity  between  true  and  predicted  daily infection ratios\footnote{{Daily infection ratio of day $t$ is defined as $r_\text{infx}(t)=(1/N)\sum_{i=1}^N y_i^{(t)}$, where $y_i^{(t)}$ is the binary infectious status of node $i$ at day $t$, $\forall\,t=1,...,T$.}}, providing a measure of how well the model fits the general trend of epidemics in the community; 
{5)~MSE: mean squared error calculated as $\frac{1}{NT} \|\bbY - \hbY \|_\mathrm{F}^2$.}
It should be noted that all metrics except AUC and MSE use the binary prediction $\bar{\bbY}$, which requires finding a classification threshold. 
We obtain that threshold as the {accuracy-}optimal cut-point in the ROC curve, calculated from the training data. 
All metrics except MSE follow a higher-is-better criterion.
It can be observed from Fig.~\ref{fig:performance_comparison} that in all test scenarios of different epidemic modeling and evolution lengths, DDmix always results in more accurate reconstruction for individual nodes (higher ACC), more distinct recognition of infected nodes apart from the healthy ones (higher AUC), more balanced predicting precision over all samples (higher F1), and a better grasp of the overall trend of epidemic evolution (higher FCS). 
In addition, DDmix is observed to be more stable than the competing methods when learning during random folds and runs. 
Since MSE is representative of the other performance metrics, we only show the MSE performance in the rest of the paper.

\vspace{2mm}\noindent
{\bf Varying the density of the graph.}
To test the generalizability of demixing approaches against topological perturbations, we vary the density of test RG graphs. 
To be specific, we fix the size $N=100$ and vary the density from the baseline $d_{r_0}=0.30$ to lower (sparser) and higher (denser) degrees with geometric modeling parameters $d_{r}^{-}=0.25$ and $d_{r}^{+}=0.35$, respectively. 
The models are trained using graphs drawn from the baseline density and are compared based on their performance on different test graphs of same, lower, and higher densities. 
For each test graph, we test the temporal reconstruction of $T\in\{10, 20\}$ consecutive steps modeled by SIRS and SIRSD, respectively. 
Table~\ref{tab:density} summarizes the models' performance using the MSE of the reconstruction as the figure of merit. 
Apart from the fact that DDmix always outperforms the baseline methods, two generic trends can be observed from all methods as follows. 
For SIRS data, the first $T{\,=\,}10$ days are easier to reconstruct than the full $T{\,=\,}20$ sequence, whereas SIRSD finds the opposite.
In the meantime, for $T{\,=\,}10$, SIRS data appears easier to reconstruct than {the more sophisticated} SIRSD, but the gap is closed in the $T{\,=\,}20$ case. 
{
Our hypothesis on this observation is that, while both the length and pattern of the epidemic can pose challenges to the learning process, the former seems to have a greater impact. 
}
Nonetheless, Table~\ref{tab:density} reveals that for all scenarios considered, DDmix proves to be significantly more accurate and robust than the baseline methods.
This observation aligns well with our hypothesis: Due to their lack of dependence on the underlying graph structure, non-graph-aware methods are unable to adjust their outputs to topologies unseen during training. 
Indeed, provided identical collapsed signals on differently structured graphs, CNN and MLP based models would always output identical reconstructions.
However, DDmix incorporates the adjacency matrix in its architecture. Thus, it is more robust to topological perturbations.\looseness=-1

\begin{table}[t]
    \captionsetup{justification=centering}
    \caption{MSE of predictions on unseen RG graphs with different graph densities for
    two different lengths of two epidemic observation models. 
    }\label{tab:density}
    \centering
    \begin{adjustbox}{width=1\linewidth}
    \begin{tabular}{c|c|r|cc|cc|cc}
        \hline
        \multicolumn{3}{r|}{Graph density} 
            & \multicolumn{2}{c|}{Baseline} & \multicolumn{2}{c|}{Denser} & \multicolumn{2}{c}{Sparser} \Tstrut\Bstrut\\\hline
        \multicolumn{3}{r|}{Time steps ($T$)}    
            & 10  & 20  & 10 & 20  & 10 & 20 \Tstrut\Bstrut\\\hline
        \parbox[t]{2mm}{\multirow{10}{*}{\rotatebox[origin=c]{90}{Epidemic model}}}
        &\parbox[t]{2mm}{\multirow{5}{*}{\rotatebox[origin=c]{90}{SIRS}}} 
        &MLP     	& .145 & .185 & .140 & .180 & .165 & .196 \Tstrut\\
        &&LSTM     	& .140 & .180 & .135 & .174 & .162 & .192 \\
        &&CNN-time    & .154 & .170 & .148 & .166 & .173 & .179 \\
        &&CNN-nodes   & .139 & .179 & .134 & .173 & .159 & .191 \\
        &&DDmix     	& \textbf{.109} & \textbf{.161} & \textbf{.115} & \textbf{.162} & \textbf{.124} & \textbf{.170} \Bstrut\\\cline{2-9}
        &\parbox[t]{2mm}{\multirow{5}{*}{\rotatebox[origin=c]{90}{SIRSD}}} 
        &MLP     	& .201 & .182 & .200 & .181 & .213 & .191 \Tstrut\\
        &&LSTM     	& .196 & .177 & .195 & .176 & .207 & .186 \\
        &&CNN-time    & .194 & .178 & .194 & .177 & .203 & .184 \\
        &&CNN-nodes   & .193 & .175 & .193 & .174 & .205 & .185 \\
        &&DDmix     	& \textbf{.165} & \textbf{.160} & \textbf{.176} & \textbf{.165} & \textbf{.172} & \textbf{.166} \Bstrut\\
        \hline        
    \end{tabular}
    \end{adjustbox}
\end{table}

\begin{figure}[t]
\centering
    \includegraphics[width=\linewidth]{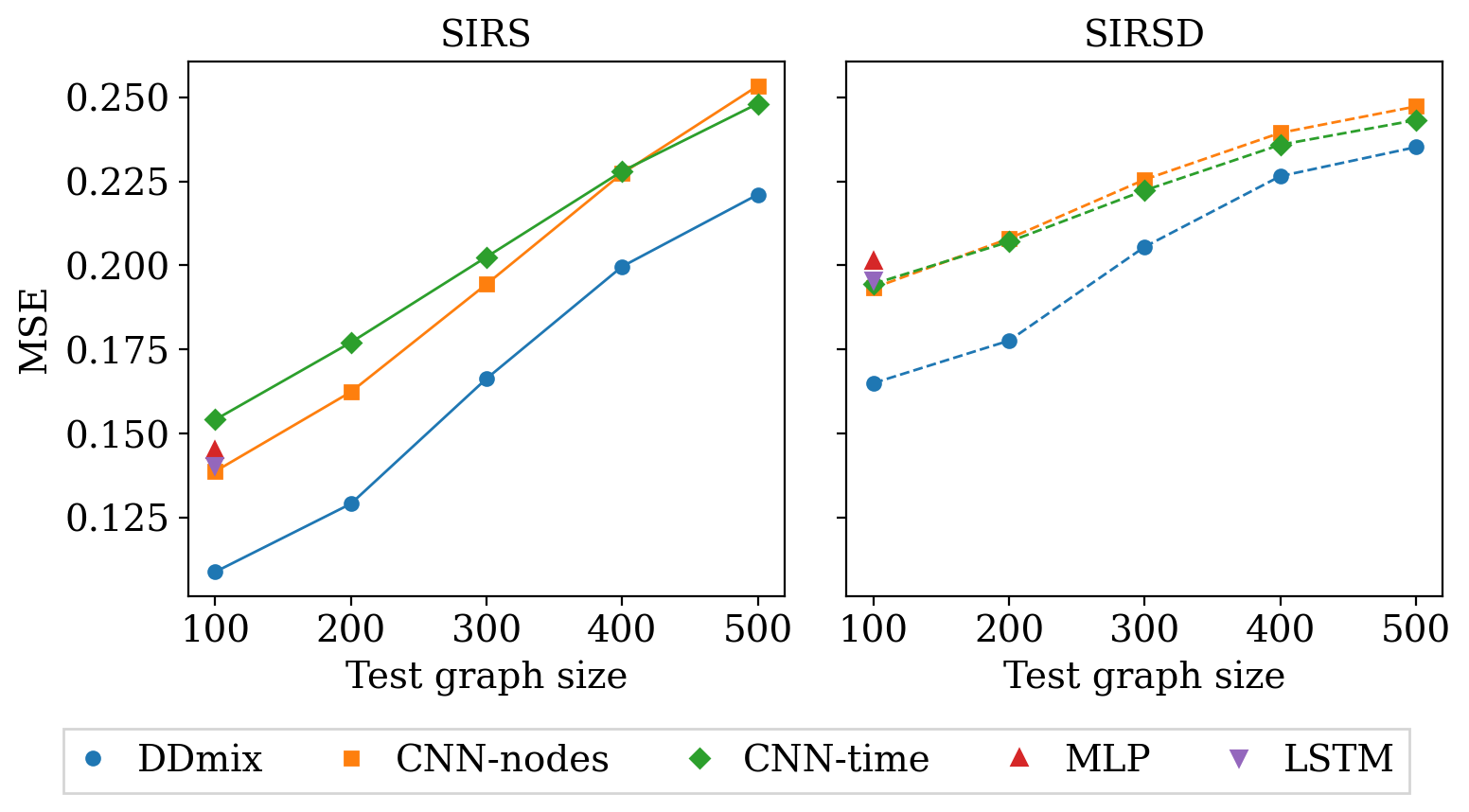}
    \caption{Comparison of models by their temporal reconstruction performance (MSE) against test graph size. 
    {The size of the training graph remains unchanged with 100 nodes.}
    On the left, we show their performance reconstructing SIRS based evolution;
    on the right, the target evolution is modeled by the SIRSD model. 
    }   
 \label{fig:size}    
\end{figure}

\vspace{2mm}\noindent
{\bf Varying the size of the graph.}
To test the generalizability of proposed methods on larger networks, we increase the size of the test RG graphs. 
We fix the number of steps to be $T{\,=\,}10$ and apply models trained on 100-node data to test graphs of size $N \in \{100, 200, 300, 400, 500 \}$. 
For generating $N$-node graphs, we set $d_r(N){\,=\,}d_{r_0}\sqrt{N_0/N}$ to control for the graph density. 
As one might expect, MLP and LSTM models are unable to handle inputs different from $N=100$. 
The variation of MSE for the other methods as a function of test graph size is depicted in Fig.~\ref{fig:size}. 
For both SIRS and SIRSD based simulations, DDmix consistently outperforms the competing approaches.

\begin{figure}[t]
\centering
    \includegraphics[width=\linewidth]{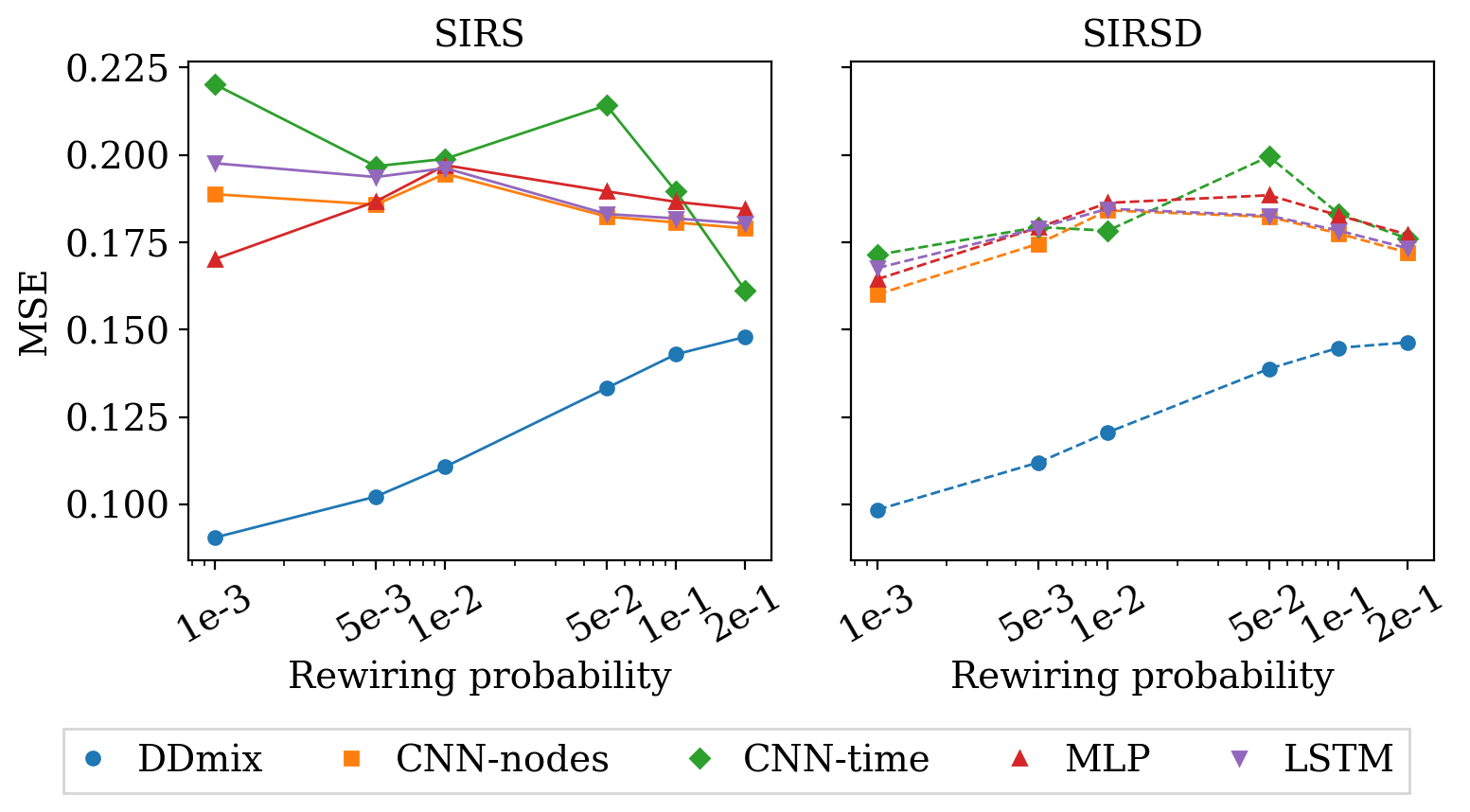}
    \caption{
    Comparison of models by their temporal reconstruction performance (MSE) against the rewiring probability of the WS graphs.
    {The models are trained and tested at each probability.}
    Both SIRS and SIRSD epidemic simulations verify the advantage of DDmix over the competing demixing methods. 
    }   
 \label{fig:rewire}    
\end{figure}

\vspace{2mm}\noindent
{\bf Varying the transitivity of the graph.}\label{sec:exp:rewire}
We extend the previous experiments to Watts-Strogatz (WS) small-world graphs~\cite{watts1998collective} to further study the impact of {\textit{transitivity}, an important topological property that measures the clustering structures in a graph by quantifying the triadic closure, i.e., how probable it is for two neighbors of a given node to be directly connected between them. 
In WS graphs, different levels of transitivity can be reflected by changing the rewiring probability $p$.}
With $p=0$, the network is highly structured as a regular ring lattice, i.e., {relatively high transitivity.
Increasing $p$ gradually removes local edges and creates short-cut paths between random pairs of nodes.
With $p$ approaching 1, the network reduces to a randomized structure close to ER graphs that yield low transitivity.}
Figure~\ref{fig:rewire} plots the MSE of temporal reconstruction approaches against the rewiring probability. 
An obvious trend can be observed from the DDmix curve that the MSE increases as a roughly logarithmic function of the rewiring probability. 
Since DDmix relies on the underlying topology, it prefers highly structured networks that have a very informative adjacency matrix; hence, it can learn useful information to accurately reconstruct the collapsed signals.
In other words, transitivity seems to be the bottleneck of DDmix's learning performance in this context.

\subsection{Reconstruction on different topologies}\label{sec:exp:graph-models}

\noindent
{\bf Training and testing on the same random graph model.}
We consider the following topology models and parameters, where all graphs have 100 nodes and the epidemics evolve for 20 steps:
\begin{enumerate}[wide,label=\roman*),topsep=0pt,itemsep=-1ex,partopsep=1ex,parsep=1ex,labelindent=0pt]
    \item {\bf BA}: Barabási–Albert graphs~\cite{albert2002statistical} that randomly grow $m=3$ edges from a new node to existing nodes;
    \item {\bf ER}: Erdős–Rényi graphs~\cite{erdos1960evolution} in which  each pair of nodes is connected with probability $p{\,=\,}0.06$;
    \item {\bf WS}: Watts-Stroga graphs~\cite{watts1998collective} with small-world property modeled by appropriate parameters $k{\,=\,}6$ and $p{\,=\,}0.005$ (see Section~\ref{sec:exp:rewire});
    \item {\bf SB}: stochastic block model graphs~\cite{holland1983stochastic} that contain three communities of sizes $\{30,30,40\}$, where edge probabilities are higher within communities ($p{\,=\,}0.170$) than between communities ($q{\,=\,}0.005$).
\end{enumerate}
We follow the same training and validation procedures as Section~\ref{sec:exp:comp-perf} and report the test results in Table~\ref{tab:graph-type}.
For each column in Table~\ref{tab:graph-type}, the models are trained and tested using data generated by the same epidemic and topology models, as the first two rows indicate. 
It can be observed that DDmix shows superior performance over the baseline methods for all epidemic and topology model combinations.
More importantly, the advantage is the most remarkable in the WS cases due to the highly structured, small-world nature.
In contrast, the other graphs, such as BA and ER, encompass much higher degrees of randomness.
This observation supports our hypothesis that DDmix is able to make good use of the information contained in the adjacency matrix.

\begin{table}[t]
    \captionsetup{justification=centering}
    \caption{
    MSE of predictions as results of learning from data evolving in different topology models. 
    }\label{tab:graph-type}
    \centering
    \begin{adjustbox}{width=1\linewidth}
    \begin{tabular}{r|cccc|cccc}
        \hline
        Epidemic 
            & \multicolumn{4}{c|}{SIRS} 
            & \multicolumn{4}{c}{SIRSD} \\\hline
        Graph  
            & BA   & ER  & WS  & SB 
            & BA   & ER  &  WS  & SB \\\hline
        MLP  
            & .184 & .188 & .187 & .194 
            & .172 & .175 & .179 & .185 \\
        LSTM 
            & .182 & .183 & .194 & .190
            & .169 & .170 & .179 & .180 \\  
        CNN-time  
            & .190 & .169 & .197 & .211
            & .186 & .174 & .179 & .185 \\
        CNN-nodes 
            & .177 & .179 & .186 & .190
            & .165 & .168 & .175 & .180 \\
        DDmix 
            & \textbf{.143} & \textbf{.147} & \textbf{.102} & \textbf{.145}
            & \textbf{.139} & \textbf{.141} & \textbf{.112} & \textbf{.145} \\
        \hline
    \end{tabular}
    \end{adjustbox}
\end{table}

\vspace{2mm}\noindent
{\bf {Inference on unseen} topology models.}
Provided that DDmix has good learning ability in various topology models, we are interested in its {out-of-distribution (OOD) performance on unseen} topology models.
More specifically, we use SB data as the training samples and, in addition, BA, ER, and WS data to test the temporal reconstruction performance of the resulting SB-based models. 
Table~\ref{tab:graph-transfer} organizes the MSE results in a similar way to Table~\ref{tab:graph-type}, except that these numbers are produced by the same models (of a certain kind) making predictions for different graphs.
For each column listed, the models reconstruct epidemic data generated by the corresponding epidemic and topology models as indicated by the first two rows.
The asterisks at the `SB' entries indicate that the model was trained using this type of graph. 
From these results, one can see that DDmix remains the best-performing demixing method. 
The performance gaps between DDmix and other methods are generally stable compared to Table~\ref{tab:graph-type} in both BA and ER cases. 
However, on the WS test data we notice a significant performance degradation in DDmix's advantage over the other methods, revealed by the average difference in MSE drops from 0.089 to 0.037 for SIRS data and from 0.066 to 0.015 for SIRSD. 
The relative worsening of DDmix is due to the lack of informative topological knowledge in the training data. 
Since WS graphs have more regular structures than SB data, the performance on WS graphs is particularly harmed by training on SB graphs. 
Nevertheless, using insufficient topological information is better than using nothing; thus, DDmix still reasonably outperforms its topology-agnostic competitors in WS graphs.

\begin{table}[t]
    \captionsetup{justification=centering}
    \caption{
    {MSE of OOD predictions}, where all models are trained using the SB graphs while tested with different topology models.
    }\label{tab:graph-transfer}
    \centering
    \begin{adjustbox}{width=1\linewidth}
    \begin{tabular}{r|cccc|cccc}
        \hline
        Epidemic 
            & \multicolumn{4}{c|}{SIRS} 
            & \multicolumn{4}{c}{SIRSD} \\\hline
        Graph  
            & BA   & ER  & WS  & SB$^*$ 
            & BA   & ER  &  WS  & SB$^*$ \\\hline
        MLP 
            & .201 & .197 & .247 & .194
            & .187 & .184 & .219 & .185 \\
        LSTM 
            & .194 & .191 & .247 & .190
            & .180 & .178 & .214 & .180 \\
        CNN-time 
            & .211 & .210 & .245 & .211
            & .187 & .185 & .204 & .185 \\
        CNN-nodes 
            & .192 & .189 & .249 & .190 
            & .179 & .177 & .216 & .180\\
        DDmix 
            & \textbf{.162} & \textbf{.159} & \textbf{.210} & \textbf{.145}
            & \textbf{.159} & \textbf{.156} & \textbf{.198} & \textbf{.145} \\
        \hline
    \end{tabular}
    \end{adjustbox}
\end{table}

\subsection{Applying DDmix to a real-world network}~\label{sec:exp:applications}

So far, we have presented extensive numerical experiments with synthetic graphs to verify the  pointwise precision of DDmix.
Now we proceed to apply this method to two applications motivated by realistic concerns and demands. 
More precisely, we leverage the demixed results to identify the source and super-spreaders of the breakout. 

\vspace{2mm}\noindent
{\bf Epidemic source tracing.}
We investigate the epidemic source identification in a real-world school network. 
Using a public dataset encoding 2 days of face-to-face contact among 232 students from 10 school classes~\cite{gemmetto2014mitigation,stehle2011high}, we construct 2 daily contact graphs: one for training ($N{\,=\,}116$) and one for testing ($N{\,=\,}228$). 
The graphs are undirected and unweighted.
An edge exists between a pair of students if they had more than 5 face-to-face contacts on that day. 
Based on these real-world graphs, 1500 SIRS simulations are generated as training samples and 1000 as test samples.
Each simulation starts from a random source node and evolves for $T{\,=\,}20$ days.
For the input node features, besides the collapsed-mean values used in the previous experiment, here we include an additional dimension of the partially available first-infection-day information. 
More precisely, it encodes on which day the infection happened for a subset of the nodes (30\% of all ever-infected nodes and never includes the actual source).

Based on this interpersonal contact network, a sourcing approach can be evaluated by its ability to correctly locate the class where the epidemic started. 
Given a reconstructed $\hbY$, we determine a top-$k$ ranking of the source class as follows. 
First, we focus on the first day for which at least one node is predicted as being infected, i.e., the smallest $t'$ for which there exists some $i$ such that $\hat{y}^{(t')}_i{\,>\,}0.5$.
We then rank the classes by the descending order of their largest prediction of infection probability.
In other words, for any class $C{\,\in\,}\ccalC$, its ranking score equals $\max_{i\in C} \hat{y}^{(t')}_i$.
The top-$k$ ranking yields $k$ classes with the highest ranking scores.
Depending on whether the true source class belongs to the top-$k$ class set, the accuracy of top-$k$ sourcing is deemed 1 or 0.
The average accuracy of DDmix is ${\{76.9,	95.4, 99.5\}}\%$ for the top-1, 3 and 5 sourcing results, respectively.
This significantly outperforms CNN-nodes' ${\{15.5, 37.5, 58.0\}}\%$ and CNN-time's ${\{10.9, 28.6, 45.5\}}\%$ mean accuracies, which are close to a random guess.\looseness=-1

\begin{figure}[t]
\centering
    \includegraphics[trim=0 0 0 0, clip, width=\linewidth]{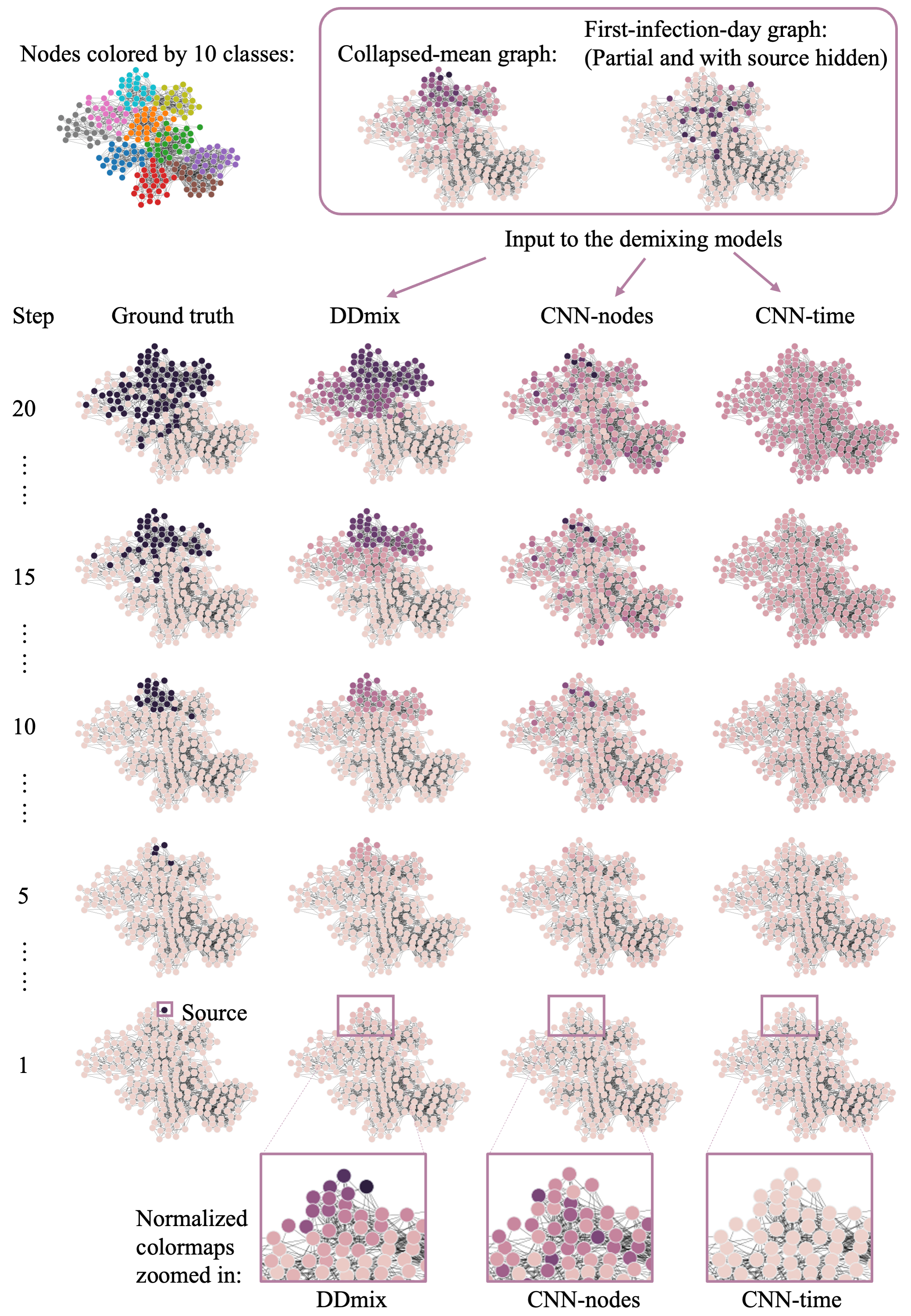}
    \caption{Visualization of one epidemic  sample showing DDmix and baseline methods tracing back to the source of infection in a real-world contact graph.}
 \label{fig:source}    
\end{figure}

Figure~\ref{fig:source} visualizes a real-world epidemic sourcing example.
The top row shows, from left to right, the contact graph categorically colored by the 10 classes, the collapsed-mean signal intensities on the contact graph, and the same graph where a small subset of nodes are marked with their first day of infection as per previous description. 
The two boxed graphs are the input $\bbx$ to all candidate demixing models, namely DDmix, CNN-nodes, and CNN-time.
Again, we note that MLP and LSTM are incompatible with different sizes of test graphs, therefore left out in this visualization. 
From a $T{\,=\,}20$ sequence, we interpolate every 5 steps and display the ground truth as well as the demixed infection scores at the selected steps.
Dark nodes denote high infection probabilities, and light nodes are predicted to be healthy. 
By default, these intensities are not processed and truthfully reflect the exact output of demixing models.
Thus, the scale of predicted probabilities for the first day may be skewed, as all but one node are healthy in the beginning.
To make the sourcing results visible, for the first day, we normalize the predicted scores by subtracting the minimum and dividing by the maximum difference. 
Based on the normalized scores, we visualize a new colormap and zoom into the neighborhood of the source.
As shown in the bottom row of Fig.~\ref{fig:source}, although the overall scale of individual probabilities is small, DDmix has successfully picked up the correct source node with the relatively highest probability.
As a comparison, CNN-nodes assigns seemingly random scores, and CNN-time assigns uniform scores to all nodes.

\vspace{2mm}\noindent
\textbf{Super-spreader assessment.}
In this experiment, we apply the proposed method to identify \textit{super-spreaders} from time-collapsed epidemic data. 
A super-spreader is an infectious node that has directly caused many infections in an epidemic spread.
To quantify the power of a spreader, we define the $s$-score of node $i$ in a $T$-step epidemic evolution as
\begin{equation}
    s(i){\,=\,}\sum\limits_{t=1}^{T-1}\sum\limits_{j\in\mathcal{N}(i)}\frac{\left[y_j^{(t+1)}-y_j^{(t)}\right]_+}{\sum_{k\in\mathcal{N}(j)}y_k^{(t)}}.
\end{equation}
Intuitively, $s(i)$ adds up the probabilities of infections spreading from node $i$ to its neighbors over time. 
The unbounded $s$-score is then normalized to $\bar{s}(i){\,=\,}2(1+e^{s(i)})^{-1}-1$ to ensure a range within 0 and 1.

With the same input data and settings as the sourcing experiment, we compute the $s$-scores of the reconstructed test samples and compare those to that of the ground truth test samples. 
Table~\ref{tab:graph-sspread} shows the MSE evaluation corresponding to each reconstruction method. 
We also provide a topology-only baseline, simply being the degree centrality of all nodes. 
It is evident that DDmix scores the best among all competitors, indicating that it is capable of attaining the most precise temporal pattern of epidemic evolution. 

\begin{table}[t]
    \captionsetup{justification=centering}
    \caption{
    MSE of normalized $s$-scores for super-spreader assessment. 
    }\label{tab:graph-sspread}
    \centering
    \begin{tabular}{r|ccccc}
        \hline
        steps & DDmix & CNN-nodes & CNN-time & Degree \\\hline
        10 & {\bf .047} & .091 & .068 & .228 \\
        20 & {\bf .094} & .137 & .195 & .109 \\\hline
    \end{tabular}
\end{table}

\section{Conclusions}\label{S:Conclusions}

We developed a novel graph CVAE architecture (DDmix) and evaluated its performance in reconstructing network epidemics from aggregated temporal observations. 
Relying on the network topology and the observation of past epidemics, DDmix adopts a graph-aware and data-driven approach to solve this challenging inverse problem of blind demixing without explicit knowledge of the epidemic model.
Numerical experiments confirmed that it outperformed non-graph-aware techniques in reconstruction accuracy and in generalizability across different graph structures, sizes, and densities. 
DDmix also emerged as a more reliable tracker of the infectious source and identifier of super-spreaders. 
Current and future research goals include: 
i)~Gearing the architecture toward the identification of key epidemic actors without the need for full temporal reconstruction, 
ii)~Investigating other observation models accounting for partially observed data (only a subset of nodes are observable) and/or temporally incomplete data (certain days of data are missing for the aggregation), and 
iii)~Analyzing how the performance and generalizability of DDmix may depend on community structures in the network topology. 

\bibliographystyle{IEEEtran}
{\footnotesize
\bibliography{citations}}

\end{document}